\documentclass[useAMS,usenatbib]{mn2e}

\usepackage{graphicx}
\usepackage{txfonts}

\title[Lyman-$\alpha$ wing absorption in cool white dwarf stars]
{Lyman-$\alpha$ wing absorption in cool white dwarf stars}

\author[R. D. Rohrmann, L. G. Althaus and S. O. Kepler]
{R. D. Rohrmann$^{1}$\thanks{E-mail: rohr@icate-conicet.gob.ar (RDR);
althaus@fcaglp.unlp.edu.ar (LGA); kepler@if.ufrgs.br (SOK)}\thanks{Member 
of the Carrera del Investigador Cient\'{\i}fico y Tecnol\'ogico, CONICET, 
Argentina.}, L. G. Althaus$^{2}$\footnotemark[1]\footnotemark[2] 
and S. O. Kepler$^{3}$\footnotemark[1]\\
$^{1}$Instituto de Ciencias Astron\'omicas, de la Tierra y del Espacio 
(CONICET), Av. Espa\~na 1512 (sur), 5400 San Juan, Argentina\\
$^{2}$Facultad de Ciencias Astron\'omicas y Geof\'isicas, UNLP, IALP-CCT 
(CONICET). Paseo del Bosque S/N, B1900FWA La Plata, Argentina\\
$^{3}$Instituto de F\'{\i}sica da UFRGS, 91501-900 Porto Alegre, RS - Brasil}
\begin{document}

\date{Accepted . Received ; in original form }

\pagerange{\pageref{firstpage}--\pageref{lastpage}} \pubyear{2010}

\maketitle

\label{firstpage}

\begin{abstract}
Kowalski \& Saumon (2006) identified the missing absorption mechanism in 
the observed spectra of cool white dwarf stars as the Ly$\alpha$ red wing 
formed by the collisions between atomic and molecular hydrogen and successfully 
explained entire spectra of many cool DA-type white dwarfs.
Owing to the important astrophysical implications of this issue,
we present here an independent assessment of the process.
For this purpose, we compute free-free quasi-molecular absorption in 
Lyman-$\alpha$ due to collisions with H and H$_2$ within the one-perturber, 
quasi-static approximation.
Line cross-sections are obtained using theoretical molecular potentials to 
describe the interaction between the radiating atom and the perturber. 
The variation of the electric-dipole transition moment with the interparticle 
distance is also considered. Six and two allowed electric dipole transitions 
due to H-H and H-H$_2$ collisions, respectively, are taken into account.
The new theoretical Lyman-$\alpha$ line profiles are then incorporated in our
stellar atmosphere program for the computation of synthetic spectra and 
colours of DA-type white dwarfs.
Illustrative model atmospheres and spectral energy distributions are computed,
which show that Ly$\alpha$ broadening by atoms and molecules has a 
significant effect on the white dwarf atmosphere models. 
The inclusion of this collision-induced opacity significantly reddens spectral 
energy distributions and affects the broadband colour indices for model
atmospheres with $T_{\mathrm{eff}}<5000$ K. These results confirm those 
previously obtained by Kowalski \& Saumon (2006).
Our study points out the need for reliable evaluations of H$_3$ potential
energy surfaces covering a large region of nuclear configurations, in order 
to obtain a better description of H-H$_2$ collisions and a more accurate 
evaluation of their influence on the spectrum of cool white dwarfs.  
\end{abstract}


\section{Introduction}

White dwarf (WD) stars  represent the most common final stage of the stellar 
evolution, and as such convey valuable information about the history of our 
Galaxy and stellar populations (see Althaus et al. 2010 for a recent review). 
The large majority of observed WDs show hydrogen-rich atmospheres (DA WDs).  
Since the spectrum of light escaping from old and cool WDs controls the rate 
at which they cool, the evaluation of cooling timescales of such WDs is 
sensitive to the gas/fluid opacity and equations of state used in the model 
atmospheres (Hansen 1998, Salaris et al. 2000, Serenelli et al. 2001). 
Furthermore, model atmospheres are a basic tool in the analysis of observed 
spectroscopy and photometric data. In this sense, the computation of synthetic 
spectra help to reveal important details about the physical processes in 
external layer of WDs and to understand the spectral evolution of these 
objects.

Photometric observations presented and analyzed by \citet{Be97} for 110 cool 
WDs showed a flux deficiency in the $B$ magnitude for stars cooler than 
$T_{\mathrm{eff}}\approx 5500$ K. There, the UV absorption was interpreted to 
be a missing opacity source due to hydrogen. By computing the Ly$\alpha$ 
quasi-molecular absorption and successful fits to the entire spectra of cool 
hydrogen atmospheres WDs, \citet{Ko06} identified the origin of $B$ deficiency 
as the Lyman-$\alpha$ line broadened by collision of absorbing H atom with 
H$_2$ molecules.

Quasi-molecular radiative transitions occur when a photon is absorbed or 
emitted by a hydrogen atom while it interacts with one or more neighboring 
particles (atom, ion or molecule) at atom-perturber separations of few 
Angstroms. The interaction between absorber atom and other particle at short
distances eventually leads to the formation of well-known molecular
satellites in the line wings \citep{St73,Sa73,Al82}.
Lyman-$\alpha$ satellites have been observed in the spectra of DA white dwarfs
and identified as produced by H+H$^+$ (1405 \AA~satellite) and H+H (1623 
\AA~satellite) collisions by \citet{Ne85} and \citet{Ko85}, respectively. 
A broad H$_2$ collision induced satellite in the red wing of Lyman-$\beta$ 
at 1150 \AA~ has been also detected by \citet{Al04} in the bright pulsating 
DA white dwarf G226-29.

Classical resonance and van-der-Waals broadening have been demonstrated to be 
inadequate \citep{Sa69} in reproducing the Lyman 
$\alpha$ red wing extending far into the optical region. These broadening 
evaluations assume that the interaction between radiating atom and perturber 
follows a power-law dependence $r^{-p}$ with the interparticle distance $r$ 
(e.g., $p=3$ for resonance broadening and $p=6$ for van-der-Waals broadening, 
c.f. Mihalas 1978), which fails to describe true interatomic potentials at 
$r$ smaller than few Angstroms where far wing contributions are produced. 
Appropriate wing absorption theories are based on quasi-molecular approaches 
which use accurate theoretical molecular potentials to describe the 
interaction between the radiator and perturber and take into account the 
variation of electric-dipole transition moment with the interparticle 
distance \citep{Ja45,Ba51,Ch57,Ga77,Al82}. 
More specifically, the spectrum of absorption is evaluated using 
{\em ab initio} calculations of Born-Oppenheimer energies and dipole transition 
moments for the electronic states of the quasi-molecule or dimer formed during 
the collision. 

Atmosphere model evaluations for cool WDs that included Lyman $\alpha$ line 
broadening based on quasi-molecule methods were performed by \citet{Ko00}, 
\citet{Wo02} and \citet{Ko06} (hereafter KS). 
KS were successful in reproducing the spectrum of cool DA-WDs
due to the inclusion of broadening by H-H$_2$. \cite{Ko00} used only 
broadening by H-H and H-He collisions.
KS evaluations were based on the semi-classical approximation which is 
generally considered valid to describe the far wings of the line profile. 
This approach takes into account the variation of the dipole moment during 
an atomic collision. \citet{Ko00}'s calculations have been made 
using the quasi-static limit of the so-called unified theory 
\citep{Al99}. In the formalism of the unified theory,  the whole line 
profile from the line core to the far wings can be assessed. It also takes 
into account additional contributions from multiple perturber 
collisions\footnote{It is not clear however how the simultaneous perturbation
of various {\em different} type of particles could be considered by 
this method.}, which are considered important at perturber densities larger 
than $10^{21}$ particles per cm$^3$. 
The so-called static limit of the unified description produces a line 
spectrum similar to that of  the semi-classical approximation.

Given that more than 80\% of the white dwarf population is of DA type
(Eisenstein et al. 2006, Althaus et al. 2010), Ly$\alpha$ wing 
broadening could have implications on the luminosity function of very 
old WDs. It is thus of much interest to study the wings of the Lyman-$\alpha$
absorption line yielded in atomic collisions. Considering KS opacity 
calculations showed that the Ly$\alpha$ quasi-molecular opacity may have 
an important role in affecting the emergent UV and blue radiation of 
cool DA-WD stars, 
we have performed an independent evaluation. KS showed only results based on 
their estimation of the main contributions to the gas opacity 
from H-H and H-H$_2$ encounters. Here, we present results including all 
dipole-allowed transitions resulting from these collisions and show an 
overall analysis of the relative importance among them. 
For computing Ly$\alpha$ wing opacity, we have elected the semi-classical 
broadening theory as a matter of convenience at the present stage of our 
calculations, to reproduce  KS calculation conditions and for 
comparison purposes in future works. We postpone to develop wing broadening 
evaluations based on the unified method to a later work.
As this paper is devoted to cool DA-type white dwarfs, we only include 
collisions of radiating H atom with atoms and molecules.

This paper is structured as follows. In Sect. \ref{s.theory} we briefly review
the semi-classical broadening method and assess at some length the wing 
broadening arising from collisions of H atoms and H$_2$ molecule. 
The line opacity including non-ideal gas effects on the upper state of 
transitions are shown in Sect. \ref{s.total}.
Model atmospheres and input physics are detailed in Sect. \ref{s.models}.
We then evaluate in Sect. \ref{s.results} the implications of our line
broadening opacity on the emergent spectrum of cool white dwarfs.
Finally, in Sect. \ref{s.conclu}, we summarize the main conclusions.


\section{Line Broadening Theory} \label{s.theory}

In the electric dipole approximation, the cross section $\sigma_\nu$ of the 
radiation absorption corresponding to a transition from an initial state $i$ 
to a final state $j$ of an atom or molecule is given by
\begin{equation} \label{e.cross}
 \sigma_{ij}(\nu) = 4\pi^2 \alpha \nu |T_{ij}|^2 \phi_{ij}(\nu),
\end{equation}
where $\alpha$ is the fine-structure constant, $\nu$ the frequency of the 
transition, $T_{ij}$ the electric-dipole transition moment, and 
$\phi_{ij}(\nu)$  the normalized line profile 
\begin{equation} \label{e.normal}
\int_0^\infty \phi_{ij}(\nu) d\nu=1.
\end{equation}
For radiation unpolarized and randomly oriented particles
$|T_{ij}|^2=\frac 13 D_{ij}(r)$, with $D_{ij}(r)$ the so-called 
dipole-strength function for the transition $i\rightarrow j$. 
As a reference, the Lyman-$\alpha$ transition ($i=1s$, $j=2p$) of isolated 
hydrogen atoms \citep{Be57}, has a central frequency 
$\nu_0=2.467\times 10^{15}$ Hz, $D_{1s,2p}=(32/27)^3a_0^2$ ($a_0$ is the 
Bohr radius) and the resulting total cross section is
\begin{equation} \label{e.isolated}
\int_0^\infty \sigma_{1s,2p}(\nu) d\nu= \frac{4\pi^2 \alpha}3 \nu_0 
D_{1s,2p} = 1.1044\times 10^{-2} cm^2.
\end{equation}

\subsection{The quasi-static approach}  

Our calculation of the wing absorption of Ly$\alpha$ is made within the
semi-classical approximation following the methodology used in KS.
This approach is based on a number of assumptions
\begin{enumerate}
  \item Born-Oppenheimer approximation: nuclei move on single adiabatic 
potential energy surfaces created by the much faster moving electrons.
  \item Adiabatic collisions: the interaction between two particles 
is viewed as the formation of a quasi-molecule which moves over a particular 
Born-Oppenheimer energy curve during the collision.
  \item Nearest neighbor approximation: the radiating atom 
is assumed to be only perturbed by its nearest neighbor.
  \item (Classical) Franck-Condon principle: the radiative transition occurs 
in the neighborhood of the internuclear distance where the difference between 
upper and lower potentials of the quasi-molecule equals the photon energy.
\end{enumerate}
The first enumerated approximation is used to identify collisionally perturbed 
atomic states as parts of the initial and final molecular adiabatic states.
The second assumption is considered valid in slow atomic collisions 
\citep{Hi54} and seems appropriate for the temperatures 
of cool WD atmospheres. \footnote{This approximation breaks down for 
high-energy collisions which can result in either ionization or excitation 
to other electronic states. Such collisions constitute non-adiabatic processes
and are considered negligible in the present work.} 
In the one-perturber approach we ignore multiple simultaneous encounters.
Classically, the fourth assumption considers that the radiative decay time 
is short compared to the collision time.

Within the preceeding assumptions, the quasi-static approximation predicts the
following expression of the line profile \citep{Ma59,Al82}
\begin{equation} \label{e.profile}
\phi_{ij}(\nu)=\frac h {|dV_{ij}/dr|} P_1(r).
\end{equation}
Here, $h$ is the Planck constant, and $V_{ij}(r)$ the difference between
the two Born-Oppenheimer energies $V_j(r)$ and $V_i(r)$ representing 
the interaction of the active atom, in each one of two states $i$ and $j$
involved in the transition, with a perturber at a distance $r$.
The first factor of the right hand side in Eq. (\ref{e.profile}) takes into
account the relationship between atom-perturber distance and transition
frequency $\nu$, which is given by
\begin{equation} \label{e.potential}
h\nu=V_{ij}(r)\equiv V_j(r)-V_i(r). 
\end{equation}
The function $P_1(r)$ in Eq. (\ref{e.profile}) is the probability 
density of finding the nearest perturber to a distance $r$ from the radiator,
\begin{equation} \label{e.probability}
P_1(r)=4\pi n_p r^2 \exp(-4\pi n_p r^3/3),
\end{equation}
being $n_p$ the mean density of perturbers in the gas.  The exponential 
factor in (\ref{e.probability}) is usually negligible numerically and is
often omitted, however, it is necessary to obtain the correct normalization 
of the profile $\phi_{ij}(\nu)$ [Eq. (\ref{e.normal})].

With Eq. (\ref{e.profile}), the cross-section in Eq. (\ref{e.cross}) may be 
written as
\begin{equation} \label{e.sec_qs}
 \sigma_{ij}(\nu) =\frac{ 4\pi^2 \alpha}3 \frac {h\nu} {|dV_{ij}/dr|} 
             D_{ij}(r) P_1(r) e^{-\beta V_i(r)},
\end{equation}
where $\beta=(k T)^{-1}$, $k$ is the Boltzmann constant and $T$ the gas 
temperature. 
The exponential factor in (\ref{e.sec_qs}) is introduced as part of the 
Boltzmann distribution function of H atoms and accounts for the probability 
of finding perturbed atoms in the lower state $i$ respect to unperturbed 
atoms [$V_i(r\rightarrow \infty)=0$]. 
If the potential difference $V_{ij}(r)$ has an irregular behavior and there
are several distances $r$ verifying the relationship (\ref{e.potential}),
we have to sum over all atom-perturber configurations 
that contribute to the wing-broadening at the frequency $\nu$.

The total cross section of a specific collision-induced absorption line is a 
sum over all lower ($i$) and upper ($j$) molecular states which can 
contribute to the wing broadening.
These contributions are additive in the Born-Oppenheimer approximation
with appropriate statistical weights $\pi_{ij}$. 
In particular, there are many contributions to the resonance broadening of 
the Lyman-$\alpha$ line corresponding to transitions of different 
quasi-molecular aggregates. 
In cool DA white dwarfs, the main perturbers are H atoms and H$_2$ 
molecules because they are the most abundant species. In the following 
sections we consider wing broadening arising from collisions of H atoms with 
these particles.

\subsection{H-H collisions} \label{s.HH} 

\begin{table}
\begin{minipage}[t]{\columnwidth}
\caption{H$_2$-states asymptotically correlated with H($1s$)+H($1s,2s,2p$).}
\label{t.states} 
\centering                      
\renewcommand{\footnoterule}{}  
\begin{tabular}{c r c c c}      
\hline\hline                    
 State & $T_e$[cm$^{-1}$] &$g_e$ & Dissociation product & Energy curve \\ 
\hline                          
 $X^1\Sigma_g^+$ &     0 & 1 & H($1s$)+H($1s$) & \footnote{\citet{Ko65}}\\
 $b^3\Sigma_u^+$ &  36113\footnote{unbound state} & 3 & H($1s$)+H($1s$) 
                 & \footnote{\citet{St99} } \\
 $B^1\Sigma_u^+$ &  90203 & 1 & H($1s$)+H($2p$) & \footnote{\citet{Wo88}}\\
 $c^3\Pi_u^+$    &  95091 & 6 & H($1s$)+H($2p$) & - \\
 $a^3\Sigma_g^+$ &  95226 & 3 & H($1s$)+H($2p$) & $^c$ \\
 $C^1\Pi_u^+$    &  99150 & 2 & H($1s$)+H($2p$) & $^a$ \\
 $E^1\Sigma_g^+$ &  99164 & 1 & H($1s$)+H($2s$) &  - \\
 $e^3\Sigma_u^+$ & 106832 & 3 & H($1s$)+H($2s$) &  - \\
 $B'^1\Sigma_u^+$& 110478 & 1 & H($1s$)+H($2s$) & $^d$ \\
 $f^3\Sigma_u^+$ & 111752 & 3 & H($1s$)+H($2p$) &  - \\
 $G^1\Sigma_g^+$ & 111812 & 1 & H($1s$)+H($2p$) &  - \\
 $h^3\Sigma_g^+$ & 112021 & 3 & H($1s$)+H($2s$) & $^c$ \\
 $I^1\Pi_g^+$    & 112072 & 2 & H($1s$)+H($2p$) &  - \\
 $i^3\Pi_g^+$    & 112216 & 6 & H($1s$)+H($2p$) & $^c$ \\
\hline                                   
\end{tabular}
\end{minipage}
\end{table}
Table~\ref{t.states} lists the products of H($1s$)-H($1s,2s,2p$) 
adiabatic collisions, including the equilibrium electronic energy $T_e$ 
for ground rotovibrational levels as given by \citet{Fi66}, 
the multiplicity $g_e$ of each state, and the data source
for energy curves relevant to the present work.

According to the adiabatic approximation, from the encounter of two H atoms, 
one in the state $1s$ and the other in the state $2p$, one of eight possible 
H$_2$ electronic states is formed (Table \ref{t.states}).
However, only four ($B^1\Sigma_u$, $a^3\Sigma_g$, $C^1\Pi_u$ and $i^3\Pi_g$) 
of these molecular levels have allowed electric dipole transitions with the 
lowest H$_2$ states, $X^1\Sigma_g^+$ and $b^3\Sigma_u^+$, which asymptotically 
correlate with two separated H($1s$) atoms.
Similarly, there are four H$_2$ levels which can be formed from adiabatic 
collisions of H($1s$) and H($2s$) atoms (Table~\ref{t.states}), but only two 
($B'^1\Sigma_u$ and $h^3\Sigma_g$) of these states have allowed dipolar 
transitions to states $X^1\Sigma_g^+$ and $b^3\Sigma_u^+$. Only states 
involved in molecular transitions allowed by electronic selection rules are 
of interest in the present study.

\begin{figure}
   \includegraphics[width=8.5cm]{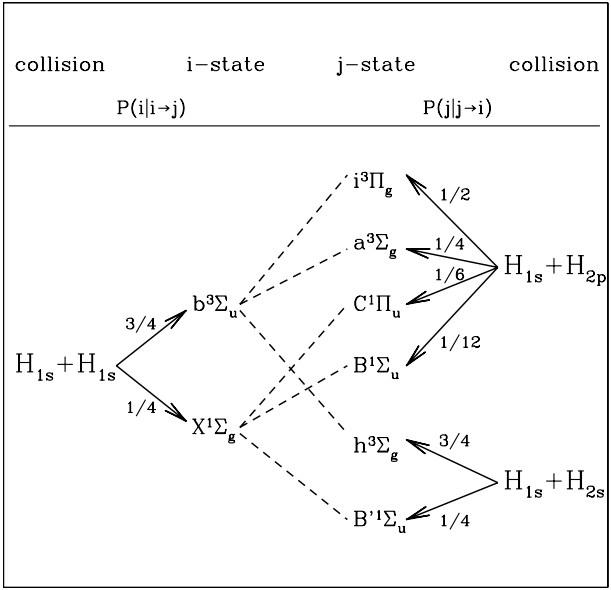} 
      \caption{Collision diagram for H($1s$)+H($1s,2s,2p$) encounters
with intermediate quasi-molecular states connected by allowed 
electric-dipole transitions (dashed lines). Conditional probabilities of 
electronic state formation (see text) are indicated on the plot. 
}
      \label{f.pi}
\end{figure}
The complete set of electric-dipole radiative processes contributing to 
Ly$\alpha$ wing from H-H collisions is illustrated in Fig. \ref{f.pi}. 
The conditional probability $P(i|i\rightarrow j)$, with values shown in Fig. 
\ref{f.pi}, is the probability that a quasi-molecule in the state $i$ is 
formed from an specific H($1s$)-H($1s,2s$ or $2p$) collision, under the 
condition that $i\rightarrow j$ is an allowed electric-dipole transition 
contributing to the Ly$\alpha$ wing (similar definition follows for 
$P(j|j\rightarrow i)$). The classical evaluation of $P(i|i\rightarrow j)$ 
takes into account that if two particles approach adiabatically each 
other, the fraction of occasions on which they move along a particular 
energy curve (says $i$) is given by the ratio of the statistical weight 
of this curve ($g_i$) to the sum of the statistical weights of all possible 
curves. 

\begin{figure}
   \includegraphics[width=8.cm]{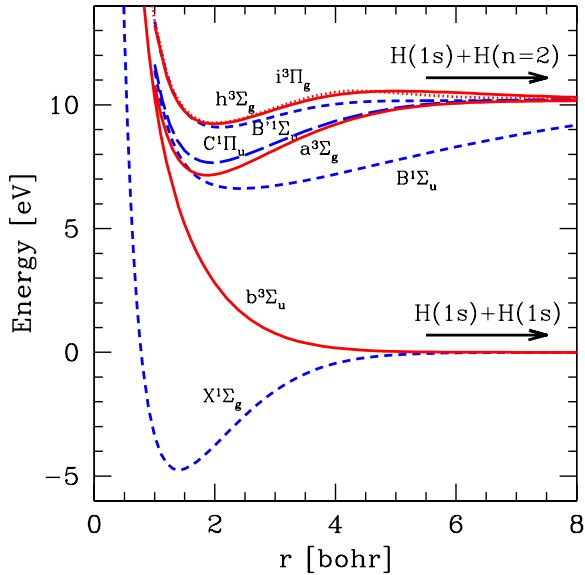} 
     \caption{Born-Oppenheimer energies of H$_2$ states relevant to this paper.
     Dashed lines are for singlet $\Sigma$ states, solid lines for 
      triplet $\Sigma$ states, long dashed line for the 
      $C^1\Pi_u^+$ state and the dotted line for the  
      $i^3\Pi_g^+$ state. The asymptotic states corresponding to
      great internuclear separations are also indicated.}
       \label{f.H2}
\end{figure}
The Born-Oppenheimer energy curves arising from H-H collisions that
contribute to the Ly$\alpha$ wings are shown in Fig.~\ref{f.H2}. The 
potential curves adopted in this work are referenced in Table \ref{t.states}.
As the atom and perturber get further away from each other, the electronic 
energies tend to asymptotic values, which are sums of individual particle 
energies. Since atomic states are degenerate, there are in general several 
molecular energy curves that tend to the same asymptotic energy.

\begin{figure}
   \includegraphics[width=8cm]{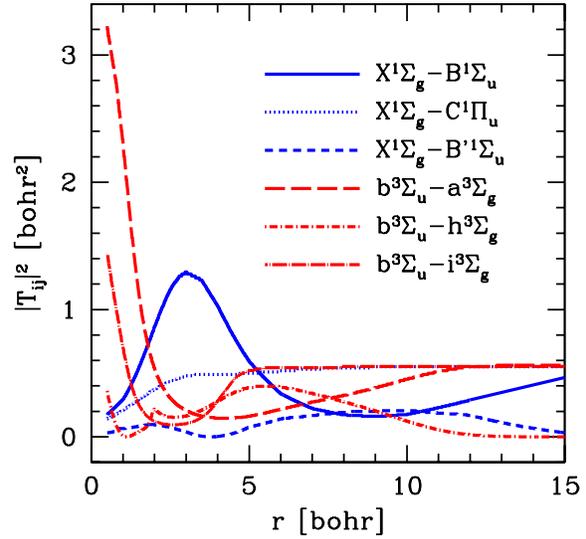} 
\caption{$|T_{ij}|^2$ dipole moments between H$_2$-states which correlates with 
H($2s$ or $2p$) $+$ H($1s$) and H($1s$) $+$ H$(1s)$ as a function of the 
internuclear distance. Data from \citet{Dr85} for $X-B$ and $X-C$, 
\citet{Fo75} for $X-B'$, and \citet{St99} for triplet state transitions.}
\label{f.moments_H2}
\end{figure}
The dipole moments of molecular transitions allowed by electronic selection 
rules are shown in Fig. \ref{f.moments_H2}.
The dipole-forbidden character of the transition ($1s-2s$) for separated 
atoms, is associated to electric-dipole transition moments of the H$_2$ 
molecule $T_{X-B'}$ and $T_{b-h}$ which become zero as the internuclear 
separations $r\rightarrow \infty$. On the other hand, the correct dipole 
transition moment for the $1s-2p$ atomic transition is obtained in the 
dissociation limit for $X-B$, $X-C$, $a-b$ and $h-b$ transitions.

To obtain the total absorption coefficient $\sigma_{H-H}(\nu)$ due to H-H 
encounters, a sum of (\ref{e.sec_qs}) over all H$_2$ transitions which 
contribute at the frequency $\nu$ must be performed,
\begin{equation} \label{e.sec_HH}
\sigma_{H-H}(\nu)= \sum_{ij} \pi_{ij} \sigma_{ij}(\nu).
\end{equation}
In the present case, the weights $\pi_{ij}$ of these contributions are 
directly related to the probabilities $P(j|j\rightarrow i)$. It can be 
demonstrated that (\ref{e.sec_HH}) recovers the correct limit value for 
asymptotically separated atoms as given by (\ref{e.isolated}).

\subsection{H-H$_2$ collisions} \label{s.HH2} 

\begin{figure}
   \includegraphics[width=8cm]{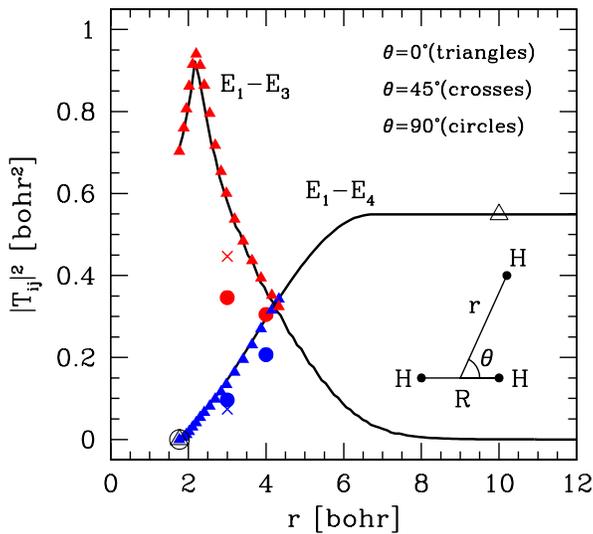} 
\caption{$|T_{ij}|$ dipole moments of $E_1-E_3$ and $E_1-E_4$ polyatomic 
transitions as a function of the nuclear geometry. 
Data from Petsalakis et al. (1988) and Peng et al. (1995) with fill and 
open symbols, respectively. Lines indicate our fits to results of Petsalakis 
et al. and Peng et al. for $\theta=0^\circ$.} 
\label{f.moments_H3}
\end{figure}

Adiabatic collisions between H atoms and H$_2$ molecules form triatomic 
systems. The analysis for atom-atom collisions in Section \ref{s.HH} was 
simplified by the fact that only a single internuclear coordinate, the 
atom-atom distance $r$, need be considered. In the present case, however, 
collision processes involving H$_2$ molecule and H atoms are functions of the 
interparticle distance $r$ and also of the molecular orientation, which can 
be characterized by a single angle $\theta$ in the case of homopolar 
diatomic molecule ($\theta$ is the angle defined between the line connecting 
the H atom to the bisector of molecular axis, see geometrical 
configuration insert in Fig. \ref{f.moments_H3}). Consequently, the 
Bohr-Oppenheimer energy solutions for the polyatomic molecule H$_3$ consist 
on energy surfaces instead of energy curves. 

Following \citet{Pe95}, we denote $E_1$, $E_2$, $E_3$ and $E_4$ 
the four lowest potential energy surfaces of the H$_3$ system. $E_1$ labels 
the ground electronic state which asymptotically separates into 
H$_2(X^1\Sigma_g)$+H($1s$).
The first excited state of H$_3$, $E_2$, adiabatically dissociates into 
H$_2(b^3\Sigma_u)$+H$(1s)$ and undergoes a conical intersection with 
the ground state at equilateral triangle ($D_{3h}$) geometries. 
$E_1$ and $E_2$ become the degenerate $2p^2E'$ state at $D_{3h}$ structures
and constitute a well-known Jahn-Teller system.
The next two excited states, $E_3$ and $E_4$ ($2s^2A'_1$ and $2p^2A''_2$ 
states in $D_{3h}$ symmetry), correlates with H$_2$($X^1\Sigma_g$)$+$H($2s$) 
and H$_2$($X^1\Sigma_g$)$+$H($2p$), respectively. These states are the lowest 
Rydberg states of H$_3$ and are located very close in energy for all 
geometries. 

It must be noted that the ($E_2$) low-lying excited state is not 
relevant to the present study because it correlates with the $b^3\Sigma_u$ 
{\em unbound} electronic state of the H$_2$-molecule rather than excited 
states of the H-atom\footnote{The $E_2$ state could be involved in ternary 
collisions since it asymptotically correlates with $3$H($1s$).}. Therefore, 
the polyatomic states of interest here are $E_1$, $E_3$ 
and $E_4$. These H$_3$ states are sufficient to infer the likely course of 
an encounter between H and H$_2$ contributing to the Ly$\alpha$ wing 
absorption (Mayne et al. 1984).

\begin{figure}
   \includegraphics[width=8cm]{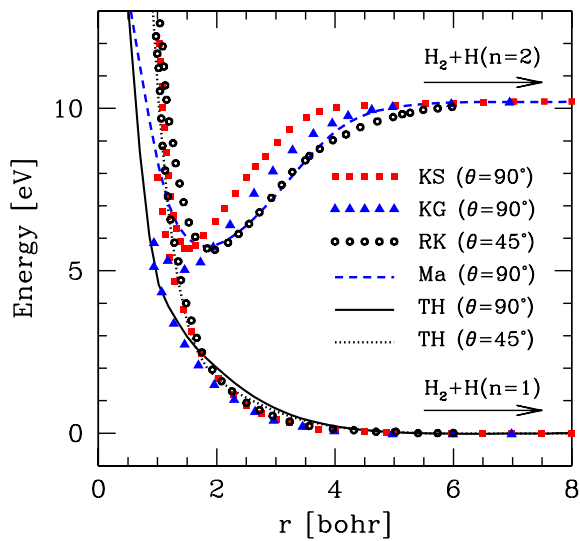} 
\caption{Electronic states of H$_3$ which in the asymptotic region H+H$_2$
correlate with H($2s$ or $2p$) $+$ H$_2(X^1\Sigma_g^+)$  and 
H($1s$) $+$ H$_2(X^1\Sigma_g^+)$ as a function of the H-H$_2$ 
interparticle distance $r$ and for collision angles $\theta=45^{\circ}$ and
$90^{\circ}$. Data from KS: \citet{Ko06} (fig. 1),  
KG: \citet{Ku79} (fig.1), \citet{Ro86} (fig. 5), Ma: \citet{Ma84} 
(analytic fit), and TH: \citet{Tr78,Tr79} (analytic fit).} 
\label{f.H3}
\end{figure}

The evaluation of potential energy surfaces (PES) is computationally 
expensive compared with similar studies of diatomic molecules. 
Theoretical calculations are often depending on the quantum 
chemical method applied and, in general, it is not easy to identify the 
best available PES, specially for excited electronic states. Moreover, 
in  the case of triatomic hydrogen, excited state energies are  often 
available for a few nuclear configurations so that interpolations or fit 
procedures are required to cover a large region of the nuclear geometry 
as it is necessary in the analysis of H-H$_2$ collisions.
Accurate energy data obtained by \citet{Li73} and by \citet{Si78} for 
the H$_3$ ground state have been fitted by \citet{Tr78,Tr79}. This fit
is considered one of the four most accurate and widely used PES for the 
H$_3$ system \citep{Mi02}. Unfortunately, energy data of the 
first Rydberg states are limited to a few geometries. 
\citet{Ku79} carried out calculations along an 
equilateral insertion path ($\theta=90^\circ$). \citet{Ro86}
used the diatomic-in-molecules (DIM) procedure to calculate the ground
and lower-lying excited PESs over several geometries, particularly for
a $45^\circ$ angle of approach.
Energy computations of \citet{Pe88} were performed mainly 
for collinear H+H$_2$ approach ($\theta=0^\circ$) and linear symmetric H$_3$,
and for a few perpendicular trajectories. More recent evaluations of Rydberg 
energies due to \citet{Pe95} contain also few data points.
A semiempirical PES of the first H$_3$ Rydberg states was developed by 
\citet{Ma84} based on evaluations in the DIM approximation \citep{Ra82}. 

Some of these energy evaluations are shown in Fig. \ref{f.H3} for non 
collinear configurations, $\theta=90^\circ$ being the more probable 
condition of impact.  Truhlar \& Horowitz (TH) analytical 
representation for the ground state agrees remarkably well 
with DIM evaluations of Roach \& Kuntz (RK) at $\theta=45^\circ$, 
and with theoretical calculations of Kulander \& Guest (KG) 
at $\theta=90^\circ$.
The equilibrium internuclear separation of the H$_2$ molecule was fixed 
at $R=1.4$ Bohr ($0.74$ \AA), except in the case of KG 
evaluations which are based on the equilibrium distance of the H$^+_3$ 
molecule ($R=1.65$ Bohr, this is only slightly larger than the bond length 
of 1.62 Bohr for H$_3$). 
Fig. \ref{f.H3} also shows values of the analytic fit due to \citet{Ma84} (Ma) 
for the first Rydberg states at $\theta=90^\circ$ 
(equivalent energy is assumed for $E_3$ and $E_4$ in this work).
Ma fit yields smaller values than those from KG calculations throughout 
the region $2<r<4.5$. At H-H$_2$ separations $\approx 2$ Bohr, the Ma curve 
crosses the KG curve and remains above of that for shorter distances. 

For comparison we have also plotted in Fig. \ref{f.H3} the results used 
by \citet{Ko06} (KS),  based on theoretical studies of \citet{Bo91,Bo96}. 
We have not incorporated results of Boothroyd et al. 
in our wing opacity evaluations because no information is available to 
which asymptotic state the evaluated levels correlate with, and there are
not enough analyzed geometrical configurations to correctly identify 
the $E_3$ and $E_4$ states from the computed data points 
(Boothroyd 2008, private communication).
The ground energy curve used by KS at $\theta=90^\circ$ lies 
above the TH and KG evaluations at H-H$_2$ separations smaller than 1.7 Bohr.
For the first Rydberg state, KS results remain above those of KG 
at distances smaller than about 5 Bohr with deviations of $\approx 0.8$ eV
for intermediate distances up to 1.5 Bohr and increasing differences at 
smaller distances. 

From this limited amount of PESs, one is already able to see the complexity 
in the selection of appropriate PESs to be used in opacity evaluations. 
Three important facts deserve to  be mentioned:
 ($i$) $r<2$ Bohr represents the critical region 
where absorptions to optical spectrum could be yielded, ($ii$) energy 
differences between $E_1$ and $E_2,E_3$ states play a decisive role in 
defining this spectrum, 
and ($iii$), among the available PES data, KG and RK results provide 
simultaneous and self-consistent evaluations of energies for these states.
Hence, we decided to adopt KG and RK data to develop PESs for the ground and 
lowest Rydberg states. In practice, we have considered the $E_3$ and 
$E_4$ potential energies identical because they are similar for all 
molecular geometries (differences smaller than $~1000$ cm$^{-1}$,
\citet{Ma84}) and because they cannot be clearly distinguished from one 
another in the available data. 
In order to construct PESs covering a large range of nuclear geometries, 
we use angular interpolation between KG ($\theta=90^\circ$) and 
RK ($\theta=45^\circ$) evaluations. In this sense, it is worth noting that 
the largest contribution to the absorption cross section comes from 
perpendicular collision paths  (the averaged-angle cross section,  
see Eq. (\ref{e.sec_HH2}) later in this Section, concentrates the 
integration weight near the configuration $\theta=90^\circ$). 
To complete the PESs, we maintained the RK values for $\theta<45^\circ$.
We have verified that this choice introduces a minor uncertainty in the cross 
section computation. The consequences for the wing absorption calculation 
due to the adopted PESs and the use of alternative PESs based on TH and Ma 
fits will be discussed in Sect. \ref{s.results}.

Although the energy surfaces for the $E_3$ and $E_4$ excited electronic states 
are nearly identical, the electric dipole moment between these states and the 
ground state differs significantly. 
In the current study, we adopt dipole moment transitions based on evaluations
of \citet{Pe88} and \citet{Pe95}. Analytical 
fits  including internuclear distances and collision angles beyond the data 
range were used in our calculations.
Fig. \ref{f.moments_H3} shows that the $|T_{ij}|$ transition moment between 
the $E_1$ and $E_4$ states increases with the atom-diatomic distance from 
$r\approx 1.8$ Bohr and approaches its theoretical value of 0.74 Bohr 
when $r\rightarrow \infty$. In $D_{3h}$ symmetry configuration this
transition is forbidden. At short interparticle distances ($r<4$ Bohr), 
the transition from the ground state to the $E_3$ state is predicted to be 
considerably stronger than the transition to the $E_4$ state. 
Although the $E_1-E_3$ transition is allowed, it decreases above $r\approx 2$
Bohr and approaches zero for $r\rightarrow \infty$, in agreement with the 
expected in the dissociation limit where the $1s\rightarrow 2s$ atomic 
transition is forbidden.

Finally, the total cross-section for Ly$\alpha$ wing broadening by H-H$_2$
collisions is averaged over all collision-frames angles and expressed as 
\begin{equation} \label{e.sec_HH2}
\sigma_{H-H_2}(\nu)= \sum_{ij} \frac{\pi_{ij}} 2
 \int_0^\pi \sigma_{ij}(\nu,\theta) \sin \theta d\theta
\end{equation}
where the summation comprises $E_1\rightarrow E_3$ and $E_1\rightarrow E_4$ 
transitions and $\sigma_{ij}(\nu,\theta)$ is given by Eq. (\ref{e.sec_qs})
with $V_{ij}(r,\theta)$, $D_{ij}(r,\theta)$ and $V_{i}(r,\theta)$ functions
of both the distance $r$ and angle orientation $\theta$ of the dimer.
The $\pi_{ij}$ weights were chosen so that, using the asymptotic values of 
H$_3$ transition moment, the isolate atom limit is recovered 
[Eq. (\ref{e.isolated})].

\subsection{Total collisional-induced wing absorption} \label{s.total}

The lines profiles $\sigma_{H-H}(\nu)$ and $\sigma_{H-H_2}(\nu)$ given
by Eqs. (\ref{e.sec_HH}) and (\ref{e.sec_HH2}) have been convolved with a 
Doppler profile to take into account the thermal broadening introduced by 
the particle motions.
We have prepared detailed opacity tabulations of these cross-sections 
appropriate for atmosphere model calculations. These results are available in 
the web site http://www.fcaglp.unlp.edu.ar/evolgroup.

The total Ly$\alpha$ wing profile $\sigma_{tot}(\nu)$ is sensitive to the 
relative abundance of the different perturbers responsible for the 
line broadening. Within the nearest neighbor quasi-static approach, the 
relative contributions to the line broadening by H and H$_2$ are given 
by the probabilities of finding an atom or molecule as the closest neighbor 
of radiating atoms. These probabilities can be approximated by the molar 
fractions of atoms ($x_H$) and molecule ($x_{H_2}$) \citep{Ro06}, 
resulting in
\begin{equation} \label{e.sec_tot}
\sigma_{tot}(\nu) = x_H \sigma_{H-H}(\nu) + x_{H_2} \sigma_{H-H_2}(\nu).
\end{equation}
Finally, the extinction coefficient (units of cm$^{-1}$) due to collisional 
induced absorptions in the Lyman$-\alpha$ line is 
\begin{equation} \label{e.sec}
 \chi(\nu) = n_{H(n=1)} \sigma_{tot}(\nu) ,
\end{equation}
where $n_{H(n=1)}$ is the number density (in cm$^{-3}$) of atoms in the 
ground state.

\section{Model atmospheres} \label{s.models}

The LTE model atmosphere code used in our analysis is a modified version
of that described at length in \citet{Ro01} and \citet{Ro02}, which is 
appropriate for hydrogen and helium atmospheric compositions (including 
mixed and pure models).
Models are computed assuming hydrostatic and radiative-convective equilibrium.
Energy transport by convection present in the cool atmospheres here
considered is treated within the usual mixing length theory, where we have
assumed the so-called ML2 parameterization of the convective flux.

The gas model used in the code includes the species H, H$_2$, H$^+$, 
H$^-$, H$^+_2$, H$^+_3$, He, He$^-$, He$^+$, He$^{2+}$, He$^+_2$, HeH and 
free electrons. The relative abundances of these species is determined by
the occupation probability formalism \citep{Hu88}.
The calculated level occupation probabilities are then explicitly included
in the calculation of the line and continuum opacities as described 
in Section \ref{s.op}.

The Ly$\alpha$ wing opacity analyzed in the preceeding sections represents a
partial contribution to the total gas opacity in atmospheres of cool DA WDs. 
For pure hydrogen models, the opacity sources in the numerical code include 
also bound-free (H, H$^-$, H$_2^+$) and free-free (H, H$_2$, H$_3$,
H$^-$, H$_2^-$, H$_2^+$) 
transitions, electronic and Rayleigh (H, H$_2$) scattering, collision-induced 
absorptions (CIA) of H-H \citep{Do68}, H-H$_2$ \citep{Gu03} and H$_2$-H$_2$ 
\citep{Bo01}, and the most significant H line series.
Details are described at long in \citet{Ro02}. It should be pointed 
out that the H$_2$-H$_2$ CIA consists on rotovibrational transitions of H$_2$ 
induced by molecular collisions, which has a strong influence on the infrared 
spectrum of cool WDs.

\subsection{Nonideal effects in opacity laws} \label{s.op}

   \begin{figure}
   \includegraphics[width=7.5cm]{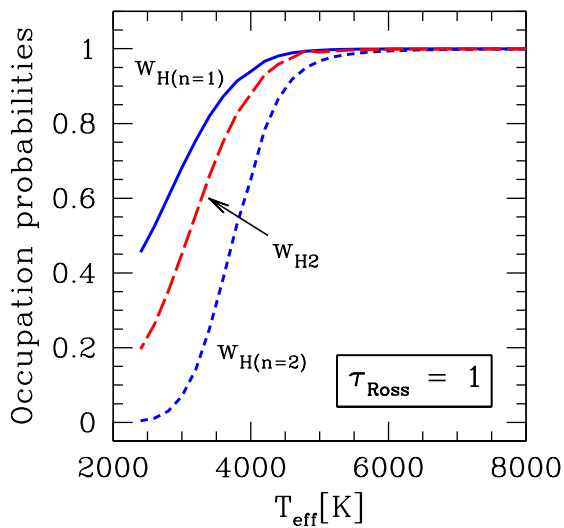} 
\caption{Occupation probabilities of the $n=1$ (dotted line) and $n=2$ 
(dashed line) hydrogen states, and of the ground H$_2$ molecule state 
(long dashed line) calculated in the photosphere ($\tau_{Ross}=1$) of 
$\log g=8$, hydrogen pure models as a function of effective temperature. 
} 
\label{f.w}
   \end{figure}

One important advance in white dwarf atmosphere modeling has been provided 
by the use of the occupation probability formalism due to \citet{Hu88} (HM). 
The HM approach considers the perturbations on each atom or 
molecule by charged and neutral particles and includes their effects in the
evaluation of atomic populations and equations of state of the gas. 
The interactions with neutral particles are treated within
the hard sphere model and those from charged particles are calculated with
micro-field distribution functions. The internal partition function of a
given specie is then written as
\begin{equation} \label{e.Z}
Z= \sum_i w_i g_i e^{-\beta Ei}
\end{equation}
where  $E_i$ and $g_i$ are, respectively, the excitation energy and 
multiplicity of the level $i$. The function $w_i$ is the so-called 
probability occupational of the level and it is computed self-consistently
with the non-ideal term in the gas free energy. The $w_i$ values 
decreases continuously and monotonically as the strength of the 
relevant interaction increases, and avoid the familiar divergence of internal
partition functions. Analytical continuity of all thermodynamical properties 
of the gas is assured by the application of the free energy minimization 
technique.
Details of our HM evaluations are given in \citet{Ro02}. 
Fig. \ref{f.w} shows the occupation probabilities of the lowest bound states 
of H and H$_2$ at a Rosseland mean optical depth $\tau_{Ross}=1$ for model 
atmospheres at 2400~K~$<T_{\mathrm{eff}}<$~8000~K and $\log g=8$.
Considering that the dominant perturbers in atmospheres of cool WDs are neutral 
particles, the occupation probabilities are mainly determined by the 
hard sphere model. The transition of $w_i$ from near unity to near zero 
is a monotonically decreasing function of the gas density which, in Fig. 
\ref{f.w}, increases with decreasing $T_{\mathrm{eff}}$.

Because the pressure shifts observed experimentally for hydrogen lines are 
very small \citep{Wi72} and because of the lack of a 
reliable theory to compute energy level shifts in bound particle states, 
the HM formalism uses energy eigenvalues of isolated particle. 
Non-ideal effects are therefore directly accounted for in $w_i$ factors
leading to the concept of an effective statistical weight for bound states, 
whereas the internal particle structure is assumed unperturbed. This includes
the use of oscillator strengths of isolated atoms.
However, since the relative population between two particle levels is 
modified respect to ideal (Boltzmann or Saha) relations [as it can be inferred 
from Eq. (\ref{e.Z})], the atomic transition rates must be accordingly 
adapted to satisfy the principle of detailed balancing at 
thermodynamic equilibrium. 
If we denote $\Lambda_{ij}$ as the transition probability per second from
$i$ to $j$ levels of an unperturbed atom, then the transition rate 
$i\rightarrow j$ for a non-ideal gas in the HM approach is given by
\citep{Hu94,Ro02}
\footnote{Notice that the proposal of transitions from bound states to 
the so-called {\em dissolved states} considered in \citet{Hu94} 
is invalid within of HM formalism. See details in \citet{Ro02}.}
\begin{equation}\label{e.L}
n_i \Lambda_{ij} w_j.
\end{equation}
This phenomenological proposal preserves the well-known Einstein and 
Einstein-Milne relations on the $\Lambda_{ij}$ coefficients. 
Thus, $\Lambda_{ij}w_j$ plays the role of a conditional probability such that 
it approaches to $\Lambda_{ij}$ (the value corresponding to isolated particles) for non-perturbed transition final states, and approaches to zero for strongly 
perturbed levels. Expression (\ref{e.L}) is applied to all bound-bound and
bound-free transitions with the convention that bare ions
have formally $w_j=1$.

The evaluation of Ly$\alpha$ wing opacity (and CIA processes in general) 
departs from the non-ideal opacity laws of the HM method. 
This happens because quasi-molecular absorptions arise from 
processes that involve atoms strongly disturbed. In fact, particle 
interactions which include variations of both energy and dipole moment of 
the transition need be directly considered in the evaluation of such
absorption cross sections \citep{Ko06b}. 
Therefore, it would be incorrect to use the non-ideal opacity law 
(\ref{e.L}) in the context of far-red wing absorption since particles 
perturbations are already explicitly included in expression (\ref{e.sec_tot}).
Of course, HM approach participates in the absorption coefficient 
(\ref{e.sec}) as it decides the actual populations of ground state atoms and 
perturbers which take part in the line wing broadening.

\section{Results} \label{s.results}

   \begin{figure}
   \includegraphics[width=8.4cm]{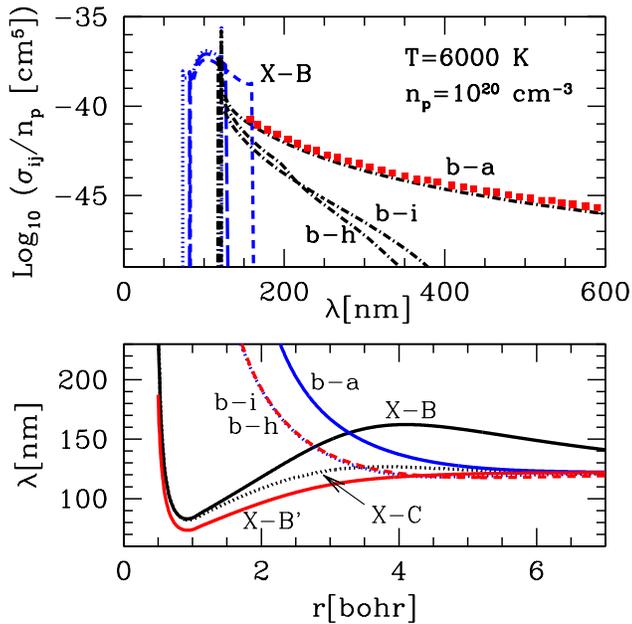}
\caption{
{\em Above:} Cross-sections as given by the quasi-static approach, Eq. 
(\ref{e.sec_qs}), for different transitions in H-H collisions. 
Dashed line for $X^1\Sigma_g^+$-$B^1\Sigma_u^+$, long dashed line for
$X^1\Sigma_g^+$-$C^1\Pi_u^+$, and dotted line for 
$X^1\Sigma_g^+$-$B'^1\Sigma_u^+$. Results due to transitions between triplet 
states are labeled on the plots (dot-dashed lines). The density of perturbers 
(H atoms) is $n_p=10^{20}$ cm$^{-3}$ and the gas temperature $T=6000$ K. 
Cross section evaluations of Kowalski (2006, fig. 49) are shown with 
symbols.
{\em Below:} Variation of the wavelength based on the level energy difference 
[Eq. (\ref{e.potential})] of selected transitions for H-H dimers as a function 
of the internuclear distance. The curves for $b-i$ and $b-h$ transitions are 
visually identical.} 
\label{f.sec_H2}
   \end{figure}
   \begin{figure}
   \includegraphics[width=8.4cm]{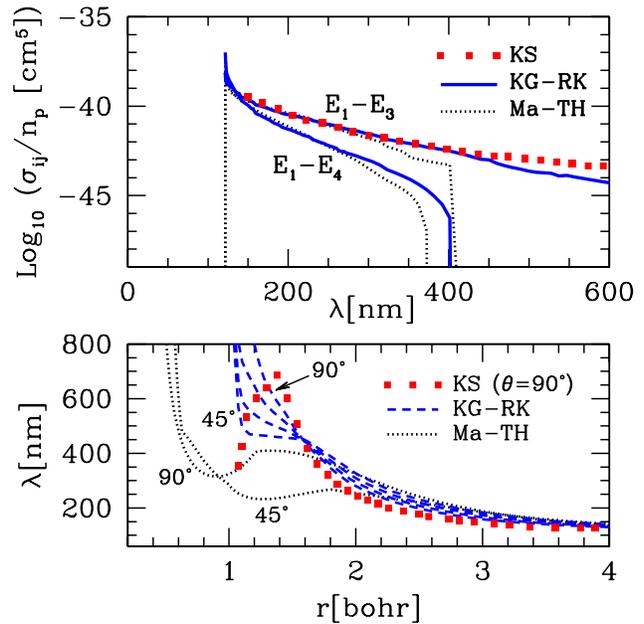} 
\caption{
{\em Above:} Cross-sections as given by the quasi-static approach, Eq. 
(\ref{e.sec_qs}), for $E_1-E_3$ and $E_1-E_4$ transitions in H$_2$-H 
collisions. Notation in the text. 
The density of perturbers (H$_2$ molecules) is $n_p=10^{20}$ cm$^{-3}$ and 
the gas temperature $T=6000$ K. 
KS results are taken from Kowalski (2006, fig. 49).
{\em Below:} Variation of the wavelength based on the level energy difference 
[Eq. (\ref{e.potential})] of select H$_3$ transitions as a function of the 
internuclear distance and for different PES evaluations
(indicated on the plot). KG-RK curves correspond to $\theta=45^\circ$, 
$60^\circ$, $75^\circ$ and $90^\circ$. KS results are based on fig. 48 
of Kowalski (2006).}
\label{f.sec_H3}
   \end{figure}

The properties of the Ly$\alpha$ broadening can be understood by 
studying Figs. \ref{f.sec_H2} and \ref{f.sec_H3}, which show the absorption
cross section of Ly$\alpha$ line broadened by collisions with H and H$_2$, 
respectively, for $T=6000$ K and perturber density $n_p=10^{20}$ cm$^{-3}$.
These figures illustrate also the photon wavelength for transitions induced 
in H-H and H-H$_2$ encounters according to the classical Franck-Condon
principle [Eq. (\ref{e.potential})].

For H-H collisions, the difference between the upper and lower interatomic 
potential for $b-a$, $b-i$ and $b-h$ triplet transitions increases 
(and the wavelength decreases) monotonically with the internuclear separation 
(bottom panel of Fig. \ref{f.sec_H2}).
Consequently, their wing spectrum contributions are relatively featureless
and extend to long wavelengths from the Ly$\alpha$ central wavelength at 
121.6 nm (top panel of Fig. \ref{f.sec_H2}). 
The potential energy difference for $X-B$, $X-C$ and $X-B'$ transitions 
does not decrease monotonically with the internuclear separation. 
Local extrema in the $X-B$, $X-C$ and $X-B'$ energy differences 
correspond respectively to wavelengths $\lambda =162.3$, 126.8, and 121.8 nm 
at the red wing, and $\lambda=83.0$, 82.0 and 73.4 nm at the blue wing.
Each extrema can eventually produce a satellite feature in the line 
wings (a satellite at 162.3 nm was predicted by \citet{Sa69}
and observed in a WD spectrum by \citet{Ko85}. These energy
extrema yield the classical discontinuities at wavelengths observed in the 
absorption cross sections for singlet transitions (Fig. \ref{f.sec_H2}). 
The blue wing due to singlet transitions is caused by the deep well in the 
ground electronic state of the diatomic molecule. In the far red wing,
the $b-a$ transition gives the dominant contribution to the line broadening
by H-H collisions. 
Comparison with results of Kowalski \& Saumon (symbols in top panel of
Fig. \ref{f.sec_H2}) shows very good agreement for the dominant absorption
at the far red wing.

\citet{Al09} noted in a recent study that simultaneous collisions with more 
than one perturber gives an additional absorption from $X-B$ transitions. 
Specifically, collisions with multiple atoms
yield $X-B$ opacity contributions which dominates the $180-300$ nm region.
These multiple collisions could introduce additional absorption 
features in the UV spectra of DA WDs, however, they have likely no impact 
over the UBVRI photometry of these stars because the corresponding bandpasses 
are at longer wavelengths. Furthermore, multiple perturber effects 
become important at densities as high as $10^{21}$ particles per cm$^3$, but 
the atomic population hardly reaches this value in WD atmospheres 
due to molecular recombination.

Top panel of Fig. \ref{f.sec_H3} shows the absorption coefficients from 
$E_1\rightarrow E_3$ and $E_1\rightarrow E_4$ transitions in H-H$_2$ collisions
as a function of the wavelength for $T=6000$ K and molecular density 
$n_{H_2}=10^{20}$ cm$^{-3}$. The variations of the wavelength with the 
H-H$_2$ distance is also given in Fig. \ref{f.sec_H3} (bottom panel).
KG-RK label denotes results based on interpolations of PES calculated by 
\citet{Ku79} and \citet{Ro86} as 
described in Sect. \ref{s.HH2}, while Ma-TH label corresponds to results
obtained from analytical fits of $E_1$ and $E_3$ ($E_4$) PESs due to
\citet{Tr78,Tr79} and \citet{Ma84}, respectively.

According to the semiclassical approximation, no broadening of the Ly$\alpha$
transition due to H$_2$ molecules will occur for $\lambda<121.6$ nm (bottom
panel in Fig. \ref{f.sec_H3}).
Cross sections based on KG-RK energy surfaces (solid lines in top panel of 
Fig. \ref{f.sec_H3}) show that the far wing broadening by H$_2$ is dominated 
by the transition $E_1-E_3$ and originated from short range interactions 
($r\la 2$ Bohr).
The $E_1-E_4$ transition gives an important contribution in the near wing
($\lambda < 150$ nm) but yields smaller cross section values than 
$E_1-E_3$ transition to longer wavelengths. As a remarkable feature, the 
cut-off at $\lambda = 400$ nm in the $E_1-E_4$ absorption is caused by the 
reduction of the dipolar moment at short internuclear distances (Fig. 
\ref{f.moments_H3}).

Fig. \ref{f.sec_H3} (bottom panel) shows that
the wavelengths derived from KS potentials at $\theta=90^\circ$ are below 
 the KG-RK values at relatively great distances ($1.5$-$4$ Bohr),
above it at intermediate distances ($1.3$-$1.5$ Bohr), and finally falls 
at short distances ($r<1.3$ Bohr). 
Such differences are relatively small in H-H$_2$ separations where the far 
wing is formed so that both calculations yield similar cross sections,
 as it can be appreciated in Fig. \ref{f.sec_H3} (top panel). Compared with 
results based on KG-RK energies, KS absorptions increase slightly towards 
long wavelengths ($\lambda > 450$ nm) likely as a result of the energy 
discrepancies found between 1.3 and 1.5 Bohr.

Fig. \ref{f.sec_H3} also shows that the opacities calculated with 
Ma-TH PESs (dotted lines) are similar to other reported results
for wavelengths shorter than $250$ nm but then fall sharply at
longer wavelengths. Discrepancies are particularly severe for the $E_1-E_3$ 
transition in the optical region ($\lambda > 400$ nm).
In this figure (bottom panel), one can see that the cross section at 
visible wavelengths is governed primarily by the energy difference 
between initial and final states in the H$_3$ transitions at 
atom-molecule separations smaller than 2 Bohr. At these distances, 
wavelengths derived from Ma-TH PESs noticeably differ from KG-RK and 
KS results. In particular, Ma-TH curves exhibit a local maximum about 
$400$ nm for $\theta=90^\circ$ (shorter wavelengths for decreasing collision
angles). Consequently, opacities based on Ma-TH energies have a dramatic 
long-wavelength reduction as shown in the top panel of Fig. \ref{f.sec_H3}.
Further analysis shows that these discrepancies are mainly due to the 
different behaviors of the PESs used for the Rydberg states.

Taking into account that absorption cross section values are very sensitive 
to small changes in the difference between the PESs of H$_3$ states, 
we are inclined to believe that KG and RK evaluations 
(which give simultaneous evaluations of $E_1$ and $E_3$ energies) provide 
a better internal consistency than the use of TH and Ma analytical fits, 
which are based on different energy computations for $E_1$ and $E_3$ 
states. A conclusion of our study is that reliable evaluations of H$_3$ 
PESs covering a large region of nuclear 
configurations are needed for a better  description of H-H$_2$ 
collisions and an accurate evaluation of their influence on Ly$\alpha$ wing 
broadening. Results shown henceforth are based on KG-RK energies.

   \begin{figure}
   \includegraphics[width=8.4cm]{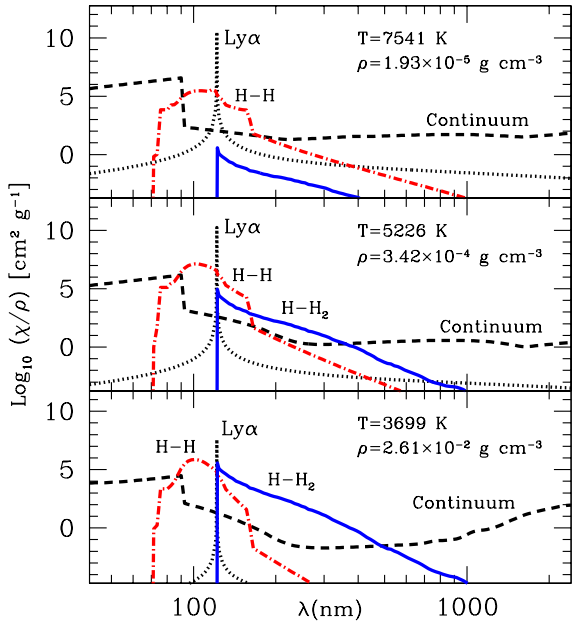} 
\caption{Monocromathic opacity coefficient in the photosphere 
($\tau_{Ross}=1$) of H model atmospheres at $\log g =8$ and 
$T_{\mathrm{eff}}=7000$ K ({\em upper}), 5000 K ({\em medium}) and 3000 K 
({\em lower}). Shown are the total continuum opacity without Ly$\alpha$ 
absorption (dashed lines), the individual contributions to Ly$\alpha$ wing 
broadening by H (dot-dashed lines) and H$_2$ (solid line), and the 
Stark-broadened Ly$\alpha$ profile (dotted lines) of \citet{Vi73}. 
Temperature and gas density values are indicated on the plot.
} 
\label{f.feos}
   \end{figure}
The importance of the collision induced broadening of Ly$\alpha$ line
can be appreciated in Fig. \ref{f.feos}, which displays the total continuous 
monochromatic opacity (dashed lines) and the Ly$\alpha$ wing broadening by H 
(dot-dashed lines) and H$_2$ (solid lines) as a function of the wavelength 
for different physical conditions of a hydrogen gas. 
These evaluations correspond to the Rosseland optical depth 
$\tau_{\rm Ross}=1$ of pure hydrogen model atmospheres at $\log g = 8$ and 
$T_{\mathrm{eff}}=7000$ K ({\em upper}), 5000 K ({\em medium}) and 3000 K 
({\em lower}). For the sake of comparison, we depict the Stark broadening 
Ly$\alpha$ profile \citep{Vi73} (dotted lines), that  accounts 
for the main source of broadening in most relatively hot WDs 
($T_{\mathrm{eff}}>10000$ K). This broadening mechanism, which is due to 
charged particle interactions, markedly weakens for the low $T_{\mathrm{eff}}$ 
analyzed here.

For the highest temperature considered in Fig. \ref{f.feos}, $T=7541$ K, 
the absorptions originated from singlet transitions in H-H interactions
dominates the gas opacity coefficient between 91 nm (the Lyman jump) 
and 160 nm. The red wing of Ly$\alpha$ is sensitive to the degree of molecular 
recombination in the gas, because that determines the relative 
importance of broadening by atoms and molecules.
At temperatures lower than $T\la 5000$ K and densities greater than 
$\rho\approx 0.0003$ g cm$^{-3}$, the degree of dissociation has been 
reduced to 40\% and the far red wing exhibits an enhanced contribution due 
to H-H$_2$ collisions. 
Below $T=4000$ K, the atoms are surrounded mostly by molecules and the 
broadening by H$_2$ dominates the total gas opacity from the Ly$\alpha$ core 
up to about 500 nm, while H-H absorption remains as the main opacity source 
at wavelengths blueward of the Ly$\alpha$ line.
These results based on KG-RK energies are in agreement with those 
obtained by \cite{Ko06}.

   \begin{figure}
   \includegraphics[width=8.4cm]{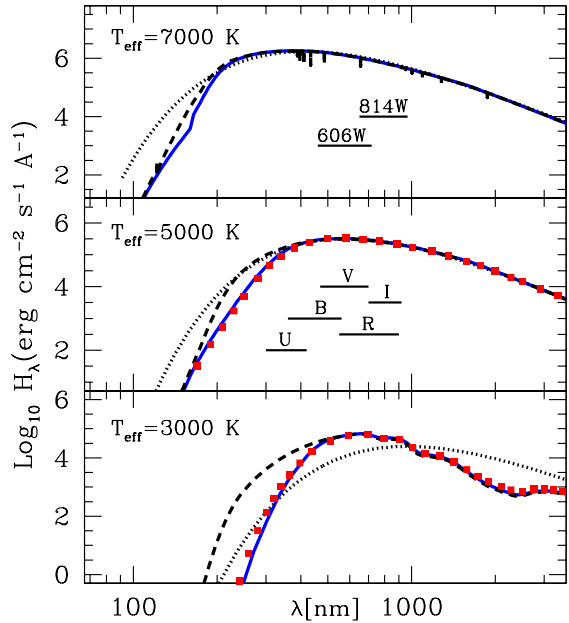} 
\caption{Emergent spectra calculated for hydrogen pure atmosphere models 
with (solid lines) and without (dashed line) collision broadening of 
Ly$\alpha$ at $\log g= 8$ and $T_{\mathrm{eff}}=7000$ K ({\em upper}), $5000$ K 
({\em medium}) and $3000$ K ({\em lower}).
Dotted lines represent black body spectra at $T=T_{\mathrm{eff}}$. 
For comparison, are shown with symbols the synthetic spectra computed 
by Kowalski (2007) for $T_{\mathrm{eff}}=3000$ and $5000$ K.
Short solid lines  indicate the location of the transmission functions for 
the filters F606W and F818W of the HST ACS (Vega-mag system) and those
of the $UBVRI$ photometry.
} 
\label{f.flux}
   \end{figure}
Synthetic spectra of DA-WD atmospheres with $\log g=8$ and 
$T_{\mathrm{eff}}=3000$, 5000 and 7000 K are show in Fig. \ref{f.flux}. 
The short wavelength spectrum of the $T_{\mathrm{eff}}=7000$ K model is 
clearly affected by H$_2$ quasi-molecular absorption between Ly$\alpha$ core 
and $\approx 200$ nm. The only prominent feature in this spectral region is 
the little jump at 160nm originated by the $X^1\Sigma_g^+-B^1\Sigma_u^+$ 
transition. At $T_{\mathrm{eff}}=5000$ K, the contribution from H$_3$ 
transitions becomes dominant far away of the line center as described above 
and significantly weakens the emergent radiation for $\lambda \la 450$ nm.
At the coolest model, $T_{\mathrm{eff}}=3000$ K, the broadening by H$_2$ 
results in a Ly$\alpha$ wing extending far into the optical region (up to 
$\approx 500$ nm).
The computed flux distribution emerges around the minimum in the
opacity caused by two collision induced absorptions, the H$_2$-H$_2$ infrared 
vibrational bands and the Ly$\alpha$ wing broadening by H-H$_2$ encounters.
Comparison in Fig. \ref{f.flux} of our new synthetic spectra (solid 
lines) with evaluations from \cite{Ko07} (symbols) for 
$T_{\mathrm{eff}}=3000$ K and $5000$ K shows an excellent agreement.

   \begin{figure}
   \includegraphics[width=8.4cm]{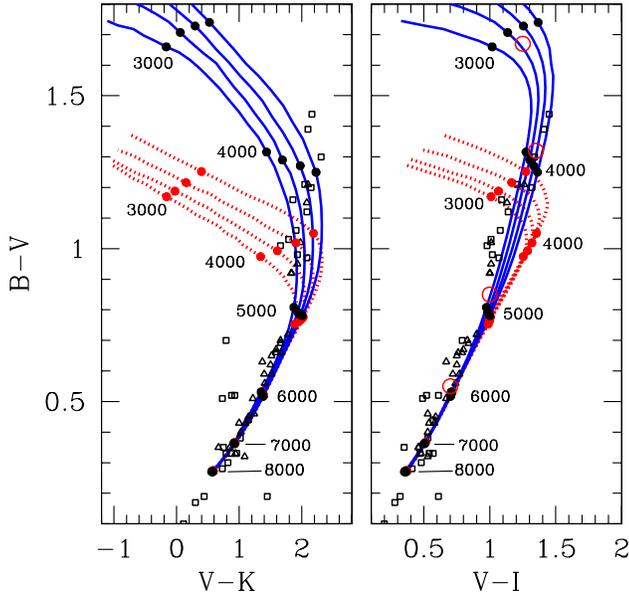} 
\caption{ [($B-V$),($V-K$)] and [($B-V$),($V-I$)] two-colour diagrams for 
hydrogen models with $\log g$ values of $7.0$, $7.5$, $8.0$ and $8.5$ 
(from top to bottom). The solid and dotted lines represent models with and 
without collision-induced wing of the Ly$\alpha$ line, respectively. 
Select $T_{\mathrm{eff}}$ values are labeled along the curves. 
Observations of DA (squares) and non-DA (triangles) WDs are taking from 
Bergeron et al. (2001). Evaluations from Kowalski \& Saumon (2006) at 
$\log g=8$ and  $T_{\mathrm{eff}}=6000$, $5000$, $4000$ and $3000$ K are 
shown as open circles.
} 
\label{f.colors}
   \end{figure}

The collision induced wing absorption does not affect significantly the 
structure of cool WD atmospheres since most of the flux is radiated at longer 
wavelengths. However, large deviations from the UV and blue colours are 
predicted with this opacity.
In Fig. \ref{f.colors} we show the computed ($B-V$, $V-K$) and ($B-V$, $V-I$) 
two-colour diagrams for our new models (solid lines) compared to observations 
(symbols) and previous model calculations without the Ly$\alpha$ red wing 
(dotted lines). These atmosphere model sequences correspond to $\log g =7$, 
$7.5$, 8 and $8.5$ (from right to left).
The effects of excluding the new opacity source are clearly visible on the
$B-V$ colour, giving rise to an increase of approximately 0.5 mag. at the 
coolest models. This additional opacity does not affect the $V$ and redder 
filters significantly. The comparison with the observed sequence of cool WDs 
\citep{Be01} shows that the new models reproduce the 
observations much better than the old models. In particular, we can see that 
the colours of the calculated models follow better the tendency of a linear 
sequence that the observations indicate. On the other hand, the sequence 
at $\log g=8$ displayed in the ($B-V$, $V-I$) diagram is very close to values 
obtained by \citet{Ko06} at select $T_{\mathrm{eff}}$ and for the same 
surface gravity.

   \begin{figure}
   \includegraphics[width=8.4cm]{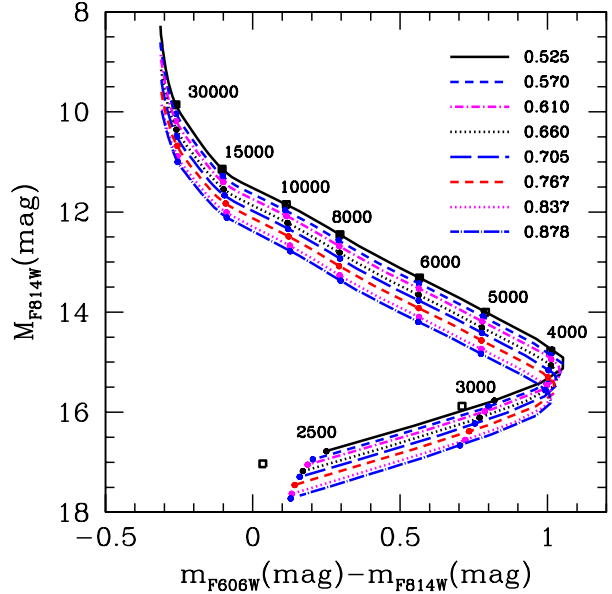} 
\caption{ ($M_{F814W},m_{F606W}-m_{F814W}$) colour-magnitude diagram 
for white dwarf sequences with masses 0.525, 0.570, 0.610, 0.660, 0.705,
0.767, 0.837 and 0.878 $M_\odot$. Pure hydrogen atmospheres are assumed.
Results for selected $T_{\mathrm{eff}}$ values are indicated with fill 
circles. Open squares correspond to the situation in which the Ly$\alpha$ 
collision broadening is not considered for the 0.525 $M_\odot$ model.
} 
\label{f.HSTcolors}
   \end{figure}
Fig. \ref{f.HSTcolors} displays a colour magnitude diagram using HST ACS
filters (Vega-mag system) for DA-WD cooling sequences of several 
stellar masses in the range $0.52<M(M_\odot)<0.88$. A similar diagram
was shown in \cite{Ko07} for $M=0.5M_\odot$. 
Present calculations are based on a homogeneous set of 
evolutionary cooling tracks of hydrogen rich DA white dwarfs \citep{Re10}. 
These models take into account the most up-to-date physical inputs including 
the complete evolutionary history of progenitor stars, element diffusion, 
chemical stratification, carbon-oxygen phase separation, and crystallization 
processes. Fig. \ref{f.HSTcolors} shows that the Ly$\alpha$ quasi-molecular 
opacity reduces the shift to the blue due to the H$_2$ collision-induced
infrared absorption in the cool extreme of the sequences.
As a result of Ly-$\alpha$ opacity, the $M_{F814W}$ magnitude and the 
$m_{F606W}-m_{F814W}$ color turn out to be 100-200 K cooler, thus implying a
larger age for a WD of a given stellar mass (see also Kowalski 2007).

\section{Conclusions} \label{s.conclu}

\cite{Ko06} (KS) identified the missing absorption in the blue and UV 
radiation of cool WDs  as due to Lyman-$\alpha$ broadening by H-H$_2$
collisions. Here, we present independent calculations for the collision 
induced broadening of Lyman $\alpha$ line in dense gases typical of cool 
white dwarf atmospheres. Our results largely agree with those of KS. 
Quasi-molecular lines in these atmospheres arise from radiative collisions of 
excited atomic hydrogen with unexcited neutral molecules H$_2$ or atoms H.
As in the KS work, present line opacity evaluations are based on the 
quasi-static approach, which relies on the nearest perturber approximation 
and the Franck-Condon principle.
Line broadening processes are therefore described in terms of independent 
absorptions occurring during collisions between a hydrogen atom and a 
perturber, either another H atom or a H$_2$ molecule. 

We revisit the properties of DA stars in the range 
8000~K~$>T_{\mathrm{eff}}>2500$~K by analyzing the synthetic spectra with 
our improved models. 
The broadband colours located in the UV and blue spectral regions are shown to 
differ substantially from those published in previous studies which did not 
include collision induced Ly$\alpha$ wing absorptions.
Full tables containing both cross-sections of collisionally-broadened 
Ly$\alpha$ wings and WD colours are available at
http://www.fcaglp.unlp.edu.ar/evolgroup or upon request to the authous at 
their e-mail addresses. 

The present study points out that a detailed knowledge of the simplest 
polyatomic molecule, H$_3$, is of fundamental importance for a precise 
understanding of Ly$\alpha$ opacity processes in cool white dwarfs. 
Accurate potential energy surfaces are required for their excited electronic 
states in order to precisely account for the effects of H-H$_2$ collisions 
on the UV and visible spectrum of these stars.

\section*{Acknowledgments}
We thank Piotr Kowalski for providing us with his computer data
for comparisons. R.D.R. acknowledges Boothroyd for valuable comments.
Part of this work was supported by the CONICET project number 
PIP 112-200801-01474 and PIP 112-200801-00940, and by AGENCIA through the 
Programa de Modernizaci\'on Tecnol\'ogica BID 1728/OC-AR.

\label{lastpage}


\begin{thebibliography}{99}

\bibitem[\protect\citeauthoryear{Allard \& Kielkopf}{1982}]{Al82} 
Allard N. F., Kielkopf J. F., 1982, Rev. of Mod. Phys., 54, 1103
\bibitem[\protect\citeauthoryear{Allard et al.}{1999}]{Al99} Allard N. F., 
Royer A., Kielkopf J. F., Feautrier, N., 1999, Phys. Rev. A, 60, 1021
\bibitem[\protect\citeauthoryear{Allard et al.}{2004}]{Al04} Allard N. F., 
H\'ebrard G., Dupuis J., Chayer P., 
Kruk J. W., Kielkopf J. F., Hubeny I., 2004, ApJ 601, L183
\bibitem[\protect\citeauthoryear{Allard \& Kielkopf}{2009}]{Al09} 
Allard N., Kielkopf J. F., 2009, A\&A, 493, 1155
\bibitem[\protect\citeauthoryear{Althaus et al.}{2010}]{Al10} Althaus L. G., 
C\'orsico A. H., Isern J., Garc\'ia-Berro  E., 2010, A\&A Rev., to be published
\bibitem[\protect\citeauthoryear{Bates}{1951}]{Ba51} 
Bates D. R., 1951, MNRAS, 111, 303
\bibitem[\protect\citeauthoryear{Bergeron et al.}{1997}]{Be97} 
Bergeron P., Leggett S. K., Ruiz M. T., 1997, ApJS, 108, 339
\bibitem[\protect\citeauthoryear{Bergeron et al.}{2001}]{Be01} 
Bergeron P., Leggett S. K., Ruiz M. T., 2001, ApJS 133, 413
\bibitem[\protect\citeauthoryear{Bethe \& Salpeter}{1957}]{Be57} 
Bethe H. A., Salpeter E. E., 1957, {\em Quantum Mechanics of One- and 
Two-Electron Atoms}, Academic Press Inc., Berlin
\bibitem[\protect\citeauthoryear{Boothroyd et al.}{1991}]{Bo91} Boothroyd A. 
I., Keogh W. J., Martin P. G., Peterson M. R., 1991, J. Chem. Phys., 95, 4343
\bibitem[\protect\citeauthoryear{Boothroyd et al.}{1996}]{Bo96} Boothroyd A. 
I., Keogh W. J., Martin P. G., Peterson M. R., 1996, J. Chem. Phys., 104, 7139
\bibitem[\protect\citeauthoryear{Borysow et al.}{2001}]{Bo01} 
Borysow A., Jorgensen U. G., Fu Y., 2001, JQSRT, 68, 235
\bibitem[\protect\citeauthoryear{Chen \& Takeo}{1957}]{Ch57} 
Chen S. Y., Takeo M., 1957, Rev. Mod. Phys., 29, 20
\bibitem[\protect\citeauthoryear{Doyle}{1968}]{Do68} 
Doyle R. O., 1968, ApJ, 153, 987
\bibitem[\protect\citeauthoryear{Dressler}{1985}]{Dr85} 
Dressler K., Wolniewicz L., 1985, J. Chem. Phys., 82, 4720
\bibitem[\protect\citeauthoryear{Eisenstein}{2006}]{Ei06} 
Eisenstein D. J., Liebert J., Harris H. C. et al., 2006, ApJS, 167, 40
\bibitem[\protect\citeauthoryear{Field et al.}{1966}]{Fi66} 
Field G. B., Sommerville W. B., Dressler K. 1966, ARA\&A, 4, 207
\bibitem[\protect\citeauthoryear{Ford et al.}{1975}]{Fo75} Ford A. L., 
Browne J. C., Shipsey E. J., DeVries P., 1975, J. Chem. Phys., 63, 362
\bibitem[\protect\citeauthoryear{Gallager \& Holstein}{1977}]{Ga77} 
Gallager A., Holstein T., 1977, Phys. Rev. A, 16, 2413
\bibitem[\protect\citeauthoryear{Gustafsson \& Frommhold}{2003}]{Gu03} 
Gustafsson M., Frommhold L., 2003, A\&A, 400, 1161
\bibitem[\protect\citeauthoryear{Hansen}{1998}]{Ha98} 
Hansen B. , 1998, Nature,  394, 860
\bibitem[\protect\citeauthoryear{Hilschfelder et al.}{1954}]{Hi54} 
Hirschfelder J. O., Curtiss C. F., Bird R. B.
{\em Molecular theory of gases and liquids} (Wiley, New York, 1954), pp. 1054
\bibitem[\protect\citeauthoryear{Hubeny et al.}{1994}]{Hu94} 
Hubeny I., Hummer D. G., Lanz T., 1994, A\&A, 282, 151
\bibitem[\protect\citeauthoryear{Hummer \& Mihalas}{1988}]{Hu88} 
Hummer D. G., Mihalas D., 1988, ApJ, 331, 794
\bibitem[\protect\citeauthoryear{e.g., Jablonski}{1945}]{Ja45} 
Jablonski A., 1945, Phys. Rev., 68, 78
\bibitem[\protect\citeauthoryear{King \& Morokuma}{1979}]{Ki79} 
King H. F., Morokuma K., 1979, J. Chem. Phys., 71, 3213
\bibitem[\protect\citeauthoryear{Kolos \& Wolniewicz}{1965}]{Ko65} 
Kolos W., Wolniewicz L., 1965, J. Chem. Phys., 43, 2429
\bibitem[\protect\citeauthoryear{Koester et al.}{1985}]{Ko85} 
Koester D., Weidemann V., Zeidler-K. T. E.-M., Vauclair G., 1985, A\&A, 142, L5
\bibitem[\protect\citeauthoryear{Koester \& Wolff}{2000}]{Ko00} 
Koester D., Wolff B., 2000, A\&A, 357, 587
\bibitem[\protect\citeauthoryear{Kowalski}{2006}]{Kk06} 
Kowalski W., 2006, Ph.D. thesis, Vanderbilt University
\bibitem[\protect\citeauthoryear{Kowalski}{2006b}]{Ko06b} 
Kowalski W., 2006b, ApJ, 651, 1120
\bibitem[\protect\citeauthoryear{Kowalski}{2007}]{Ko07} 
Kowalski W., 2007, A\&A, 474, 491
\bibitem[\protect\citeauthoryear{Kowalski \& Saumon}{2006}]{Ko06} 
Kowalski W., Saumon D., 2006, ApJ, 651, L137 (KS)
\bibitem[\protect\citeauthoryear{Kulander \& Guest}{1979}]{Ku79} 
Kulander K. C., Guest M. F., 1979, J. Phys. B, 12, L501 (KG)
\bibitem[\protect\citeauthoryear{Liu}{1973}]{Li73} 
Liu B., 1973, J. Chem. Phys., 58, 1925
\bibitem[\protect\citeauthoryear{Margeneau \& Lewis}{1959}]{Ma59} 
Margeneau H., Lewis M., 1959, Rev. of Mod. Phys., 31, 569
\bibitem[\protect\citeauthoryear{Mayne et al.}{1984}]{Ma84} 
Mayne H. R., Polanyi J. C., Sathyamurthy N., Raynor S., 1984, 
J. Phys. Chem., 88, 4064 (Ma)
\bibitem[\protect\citeauthoryear{Mielke et al.}{2002}]{Mi02} 
Mielke S. L., Garret B. C., Peterson K. A., 2002, J. Chem. Phys., 116, 4142
\bibitem[\protect\citeauthoryear{Mihalas}{1978}]{Mi78} 
Mihalas D., 1978, {\em Stellar Atmospheres}, 2nd edn., Freeman, San Francisco
\bibitem[\protect\citeauthoryear{Nelan \& Wegner}{1985}]{Ne85} 
Nelan E. P., Wegner G., 1985, ApJ, 289, L31
\bibitem[\protect\citeauthoryear{Peng et al.}{1995}]{Pe95} 
Peng Z., Kristyan S., Kuppermann A., 1995, Phys. Rev. A, 52, 1005 
\bibitem[\protect\citeauthoryear{Petsalakis et al.}{1988}]{Pe88} 
Petsalakis I., Theodorakopoulos J., Wright J. S., 1988,
J. Chem. Phys., 89, 6850
\bibitem[\protect\citeauthoryear{Raynor \& Herschbach}{1982}]{Ra82} 
Raynor S., Herschbach D. R., 1982, J. Chem. Phys., 86, 1214
\bibitem[\protect\citeauthoryear{Renedo et al.}{2010}]{Re10} 
Renedo I., Althaus L.G., Miller Bertolami M. M., Romero A. D., 
C\'orsico A. H., Rohrmann R. D., Garc\'ia-Berro E., 2010, ApJ, 717, 183 
\bibitem[\protect\citeauthoryear{Roach \& Kuntz}{1986}]{Ro86} 
Roach A. C., Kuntz P. J., 1986, J. Chem. Phys., 84, 822 (RK)
\bibitem[\protect\citeauthoryear{Rohrmann}{2001}]{Ro01} 
Rohrmann R. D., 2001, MNRAS, 323, 699
\bibitem[\protect\citeauthoryear{Rohrmann et al.}{2002}]{Ro02} 
Rohrmann R. D., Serenelli A. M., Althaus L. G., Benvenuto O. G., 2002, 
MNRAS, 335, 499
\bibitem[\protect\citeauthoryear{e.g. Rohrmann \& Zorec}{2006}]{Ro06} 
Rohrmann R. D., Zorec J., 2006, Phys. Rev. E, 74, 041120
\bibitem[\protect\citeauthoryear{Salaris et al.}{1969}]{Sa00} 
Salaris M., Garc\'ia-Berro E., Hernanz M., Isern J., Saumon D., 2000,
ApJ, 544, 1036
\bibitem[\protect\citeauthoryear{Sando et al.}{1969}]{Sa69} 
Sando K., Doyle R. O., Dalgarno A., 1969, ApJ, 157, L143
\bibitem[\protect\citeauthoryear{Sando \& Wormhoudt}{1973}]{Sa73} 
Sando K., Wormhoudt J. C., 1973, Phys. Rev. A, 7, 1889
\bibitem[\protect\citeauthoryear{Serenelli et al.}{2001}]{Se01} 
Serenelli A. M., Rohrmann R. D., Althaus L. G., Benvenuto O. G., 2001, 
MNRAS, 325, 607
\bibitem[\protect\citeauthoryear{Siegbahn \& Liu}{1978}]{Si78} 
Siegbahn P., Liu B., 1978, J. Chem. Phys., 68, 2457
\bibitem[\protect\citeauthoryear{Staszewska \& Wolniewicz}{1999}]{St99} 
Staszewska G., Wolniewicz L., 1999, J. Mol. Spectr., 198, 416
\bibitem[\protect\citeauthoryear{Stewart et al.}{1973}]{St73} 
Stewart J. C., Peek J. M., Cooper J., 1973, ApJ, 179, 983
\bibitem[\protect\citeauthoryear{Truhlar \& Horowitz}{1978}]{Tr78} 
Truhlar D.G., Horowitz C.J., 1978, J. Chem. Phys., 68, 2466 (TH)
\bibitem[\protect\citeauthoryear{Truhlar \& Horowitz}{1979}]{Tr79} 
Truhlar D.G., Horowitz C.J., 1979, J. Chem. Phys., 71, 1514 
\bibitem[\protect\citeauthoryear{Vidal et al.}{1973}]{Vi73} Vidal C. R., Cooper J., Smith E. W., 1973,
ApJS, 25, 37
\bibitem[\protect\citeauthoryear{Wiese et al.}{1972}]{Wi72} 
Wiese W. L., Kelleher D. E., Paquette D. R., 1972, Phys. Rev., A 6, 1132
\bibitem[\protect\citeauthoryear{Wolff et al.}{2002}]{Wo02} 
Wolff B., Koester D., Liebert J., 2002, A\&A, 385, 995
\bibitem[\protect\citeauthoryear{Wolniewicz \& Dressler}{1988}]{Wo88} 
Wolniewicz L., Dressler K., 1988, J. Chem. Phys., 88, 3861 

\end{thebibliography}
\end{document}